\documentclass[11pt,a4paper]{article}
\usepackage[margin=28mm]{geometry}
\usepackage[T1]{fontenc}
\usepackage{mathptmx}
\usepackage{microtype}
\usepackage{amsmath,amssymb,amsthm}
\usepackage{booktabs,array}
\usepackage{natbib}
\usepackage{tikz}
\usetikzlibrary{calc}
\usepackage{caption,subcaption}
\usepackage{fancyhdr}
\usepackage{xcolor}
\definecolor{linkblue}{RGB}{28,53,73}
\definecolor{faceblue}{RGB}{209,224,234}
\definecolor{lineblue}{RGB}{54,84,105}
\usepackage[colorlinks=true,allcolors=linkblue]{hyperref}
\hypersetup{pdftitle={Robust Downward Reasoning and Rationalizability},pdfauthor={Shuige Liu and Gabriel Ziegler}}
\newtheorem{theorem}{Theorem}
\newtheorem{proposition}{Proposition}
\newtheorem{lemma}{Lemma}
\newcommand{\RP}{\mathrm{RP}}
\tikzset{edge/.style={draw=lineblue,line width=.7pt},face/.style={fill=faceblue,draw=lineblue,line width=.7pt},vertex/.style={circle,fill=lineblue,inner sep=1.6pt}}

\title{Robust Downward Reasoning and Rationalizability\thanks{The text and formal results were prepared with Chat GPT-6 Astra; the model setup and research question are the authors'. The authors take full responsibility for the content.}}
\author{\begin{tabular}[t]{c}Shuige Liu\\[3pt]\small Center for Mathematical Economics (IMW)\\\small Bielefeld University\\\small \texttt{shuige.liu@uni-bielefeld.de}\end{tabular}\hspace{.7cm}\begin{tabular}[t]{c}Gabriel Ziegler\\[3pt]\small Freie Universit\"at Berlin\\\small Berlin School of Economics\\\small \texttt{gabriel.ziegler@fu-berlin.de}\end{tabular}}
\date{\today}

\begin{document}
\maketitle
\thispagestyle{empty}
\begin{abstract}
We study whether downward reasoning can exclude a rationalizable action at every level above 1 under every level-0 anchor. By downward reasoning, we mean that, for each fixed anchor, a player above level 1 may best respond to any mixture of opponent actions from lower positive levels. We show that such an exclusion requires at least four actions for each player and at least nine actions in total. A $5\times4$ game attains these bounds. The lower bound follows from the geometry of simultaneous best replies and a strict-dominance result. We represent sets of actions that can be best replies to the same conjecture by a simplicial complex. Its topological properties, namely connectedness and an Euler characteristic of one, restrict the possible configurations of best-reply sets in small games.
\end{abstract}
\noindent\textit{Keywords:} rationalizability, downward reasoning, best-reply sets, simplicial complexes, nerve theorem.\\
\textit{JEL classification:} C72, D83.

\section{Introduction}

In this note, we ask how small a game can be if rationalizability permits an action that robust downward reasoning excludes. We consider a positive-level version of the reasoning studied by \citet{LiuZiegler2026}. The anchor specifies a mixed action for each player at level 0.
At level 1, each player may choose any best reply to the opponent's level-0 play.
At every higher level, a player may best respond to any mixture of opponent actions from lower \emph{positive} levels.
We allow every anchor and hold it fixed throughout each reasoning process. The question is whether a rationalizable action can remain excluded at every level above 1 under every such starting point.\footnote{The restriction to lower positive levels is essential for our question.
If players could also best respond to level-0 play at higher levels, as under downward rationalizability in \citet{LiuZiegler2026}, every rationalizable action could arise at every positive level under some anchor.
We thus consider a strengthening of downward rationalizability that permits a separation from rationalizability while leaving conjectures over lower positive levels unrestricted.}

We show that the smallest example has nine actions in total. Each player must have at least four actions, and no $4\times4$ game has the required exclusion. A $5\times4$ game does. In that game every action is rationalizable, but two actions of the player with five actions are excluded at every level above 1. The example prevents the cumulative reasoning process from bringing together two opponent actions that are jointly needed to justify either excluded action.

Level-based models allow strategic behavior to depend on a player's assessment of the reasoning of others (see \citet{StahlWilson1995}, \citet{Nagel1995}, and \citet{CamererHoChong2004}). \citet{BrandenburgerFriedenbergKneeland2025} give a $4\times4$ game in which every action is rationalizable, but some actions cannot arise at any classic level above 1, for any anchor. Their example concerns responses to the immediately preceding level. Allowing mixtures of all lower positive levels changes the comparison, as our lower bound shows. \citet{HalevyHoelzemannKneeland2025} study experimentally how behavior departs from restrictions imposed by iterative top-down reasoning. Our concern is the smallest game in which the robust positive-level solution excludes a rationalizable action. We focus on solution concepts throughout.\footnote{For the foundations of downward reasoning and its relation to other level-based models, see \citet{LiuZiegler2026}.}

The proof uses a simple representation of simultaneous best replies.
For each action, consider the convex set of conjectures that justify it.
Intersections of these sets describe which actions can be optimal together.
We represent this information by a simplicial complex whose vertices are actions, whose edges join pairs of actions that can be optimal together, and whose higher-dimensional faces represent larger sets of simultaneous best replies.
Because the best-reply regions cover the simplex of conjectures and their nonempty intersections are convex, the resulting complex is connected and has Euler characteristic one.
The latter property imposes a restriction on the number of vertices, edges, triangles, and higher-dimensional faces.
With at most four actions, these restrictions leave few possibilities.
A result of \citet{Weinstein2020} rules out the remaining configuration.
Geometric representations of best replies also appear in the
literature on equilibrium computation \citep{vonStengel2002,Egan2008}
and admissibility \citep{BrandenburgerFriedenbergKeisler2008}.
While \citet{BrandenburgerFriedenbergKeisler2008} study faces of the
polytope of attainable payoff vectors, we study intersections of
best-reply regions in the simplex of conjectures. To the best of our knowledge, we are the first to use the
topological properties of simultaneous best-reply sets to
derive restrictions on level-based reasoning.

\section{Solution concepts and the bound}

Consider a finite two-player static game.
Player $i\in\{1,2\}$ has a nonempty action set $A_i$ and payoff function $u_i$.
Write $-i$ for the other player and extend payoffs linearly to mixed actions and conjectures.
For a conjecture $\mu\in\Delta(A_{-i})$ about the opponent's action, let $r_i(\mu)$ be the set of all best replies in $A_i$.
For a nonempty $S_{-i}\subseteq A_{-i}$, define
\[
\rho_i(S_{-i})=\bigcup_{\mu\in\Delta(S_{-i})}r_i(\mu).
\]
Thus, an action belongs to $\rho_i(S_{-i})$ if and only if it
is a best reply to some conjecture supported on $S_{-i}$.
We then say that the action is justified by $S_{-i}$.
Rationalizability starts from $R_i^0=A_i$ and successively requires best replies to surviving opponent actions:
\[
(\forall n\geq0)\quad R_i^{n+1}=\rho_i(R_{-i}^n)
\quad\text{ and }\quad
R_i^\infty=\bigcap_{n\geq0}R_i^n.
\]
In a finite game the iteration stop at some finite order. We use the standard best-reply characterization of rationalizability \citep[Chapter 4]{BattigalliCatoniniDeVito2025}.

A level-0 anchor assigns a mixed action to each player.
Write $\mu_i\in\Delta(A_{-i})$ for the mixed action assigned to player $-i$.
Player $i$'s level-1 set is $E_i=r_i(\mu_i)$, the set of all best replies to $\mu_i$, including ties.
Following \citet{Weinstein2020}, we call a set of this form an exact best-reply set.
Write $E=E_1\times E_2$ for the resulting product. Allowing every anchor gives the family
\[
\mathcal E=\bigl\{r_1(\mu_1)\times r_2(\mu_2):
(\mu_1,\mu_2)\in\Delta(A_2)\times\Delta(A_1)\bigr\}.
\]
The recursion depends on the anchor only through $E$.
For $E\in\mathcal E$, let $\tau_i^k(E)$ collect player $i$'s actions at the first $k$ positive levels:
\[
\tau^1(E)=E
\quad\text{and}\quad
(\forall k\geq1)\quad
\tau_i^{k+1}(E)=\tau_i^k(E)\cup\rho_i(\tau_{-i}^k(E)).
\]

At level $k\geq2$, player $i$ can choose any action in
$\rho_i(\tau_{-i}^{k-1}(E))$.
We hold the anchor, and hence $E$, fixed throughout the recursion.
Since the cumulative sets increase and the game is finite, the
recursion stops after finitely many iterations at a product
$\tau^\infty(E)$.
By construction,
\[
\forall i\in\{1,2\}\quad
\rho_i(\tau_{-i}^\infty(E))\subseteq\tau_i^\infty(E).
\]
Thus, $\tau^\infty(E)$ is closed under rational behavior (CURB).\footnote{A nonempty product $S_1\times S_2$ is closed under rational behavior if $\rho_i(S_{-i})\subseteq S_i$ for both players; see \citet{BasuWeibull1991}.}
By monotonicity of $\rho_i$, every CURB product containing $E$
also contains $\tau^k(E)$ at every iteration.
Hence, $\tau^\infty(E)$ is the smallest CURB product containing $E$.

Write $\RP_i$ for the actions available at \emph{some} level $k\geq2$ under some anchor.
Finiteness and monotonicity give
\[
\RP_i=\bigcup_{E\in\mathcal E}\rho_i(\tau_{-i}^\infty(E)).
\]
We compute the recursion for each $E$ before taking the union.
An action in $E_i$ remains in $\tau_i^\infty(E)$ even if it never appears above level 1.
The set $\rho_i(\tau_{-i}^\infty(E))$ consists of the actions available at some level above 1 for that $E$.\footnote{For any conjecture $\mu$ over the opponent's actions, we can choose $\mu_i=\mu$ as the opponent's level-0 anchor.
Every best reply to $\mu_i$ is then available at level 1 under the resulting anchor.
The level-1 sets alone thus cover $R_i^1=\rho_i(A_{-i})$ as the anchor varies.
Counting level 1 in $\RP_i$ would therefore include every rationalizable action.}

\begin{theorem}\label{thm:minimum}
If $R_i^\infty\setminus\RP_i$ is nonempty for either player, then
\[
(\forall i\in\{1,2\})\quad |A_i|\geq4
\quad\text{ and }\quad |A_1|+|A_2|\geq9.
\]
There is a $5\times4$ game in which all actions are rationalizable and $R_1^\infty\setminus\RP_1$ is nonempty.\footnote{In general, $\RP_i\subseteq R_i^2$, but
$\RP_i$ need not be contained in $R_i^\infty$.
Indeed, $\tau_i^\infty(E)\subseteq R_i^1$ for every $E\in\mathcal E$,
so the former inclusion follows from monotonicity of $\rho_i$.
For the latter observation, consider the game
\[
\begin{array}{c|cc}
  & c & d\\ \hline
x & 0,1 & 2,0\\
y & 2,1 & 0,0\\
z & -1,0 & -1,1
\end{array}
\]
The unique rationalizable action profile is $(y,c)$.
Under an anchor assigning probability one to $z$ for player 1,
player 2 chooses $d$ at level 1.
Player 1 can then choose $x$ at level 2.
Thus, $x\in\RP_1\setminus R_1^\infty$. The example is minimal.
If either player has only one action, the rationalizability
procedure stabilizes after one round.
In a $2\times2$ game, it stabilizes after at most two rounds.
Since $\RP_i\subseteq R_i^2$, no smaller game can exhibit
this phenomenon.}
\end{theorem}

The minimum total number of actions is therefore nine. Up to interchanging players, $(5,4)$ is also the unique coordinatewise minimal size. We first give the example and then prove the lower bound.

\section{A game with nine actions}

Player 1 has actions $A_1=\{e,f,x,y,h\}$ and player 2 has actions $A_2=\{a,b,c,d\}$. Payoffs, with player 1's payoff first, are
\[
\begin{array}{c@{\quad}rrrr}
\toprule
 &a&b&c&d\\
\midrule
e&-1,4&-1,0&2,3&3,0\\
f&-1,4&-1,0&3,0&2,3\\
x&0,0&0,4&4,3&0,0\\
y&0,0&0,4&0,0&4,3\\
h&2,4&2,0&2,0&2,0\\
\bottomrule
\end{array}
\]

\begin{proposition}\label{prop:example}
In this game,
\[
(\forall i\in\{1,2\})\quad R_i^\infty=A_i
\quad\text{ and }\quad \RP_1=\{x,y,h\}.
\]
\end{proposition}

\begin{proof}
Write $s$ and $t$ for the probabilities of $c$ and $d$. Abusing notation, Player 1's expected payoffs are
\[
\begin{gathered}
u_1(e)=3s+4t-1\quad\text{ and }\quad u_1(f)=4s+3t-1,\\
u_1(x)=4s\quad\text{ and }\quad u_1(y)=4t,\\
u_1(h)=2.
\end{gathered}
\]
Figure~\ref{fig:br-regions} depicts player 1's best-reply
regions as the probabilities $s$ and $t$ vary.
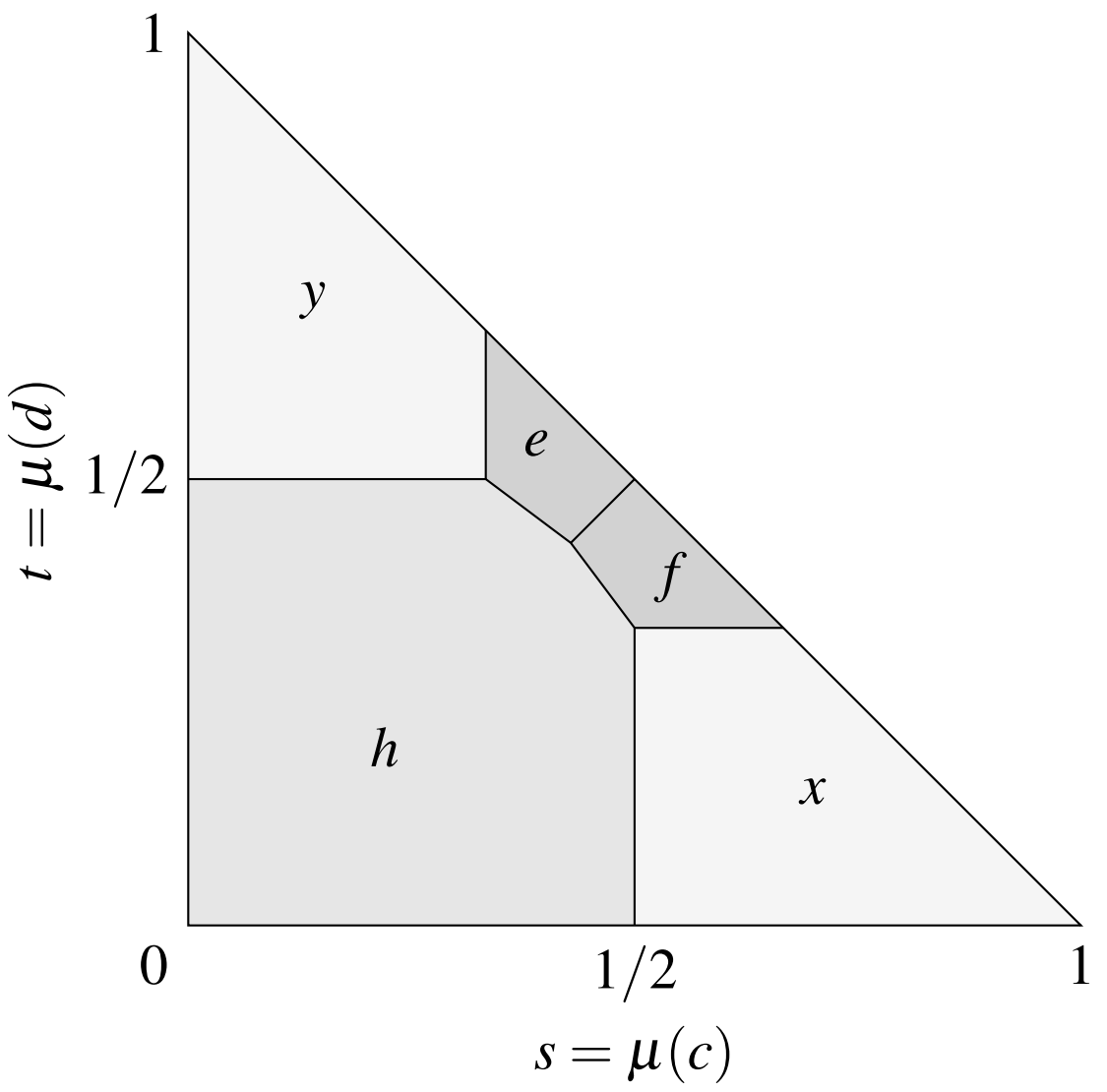
\begin{figure}[htbp]
\centering
\begin{tikzpicture}[scale=6]

\fill[gray!20]
    (0,0)--(.5,0)--(.5,{1/3})
    --({3/7},{3/7})--({1/3},.5)
    --(0,.5)--cycle;

\fill[gray!8]
    (.5,0)--(1,0)--({2/3},{1/3})
    --(.5,{1/3})--cycle;

\fill[gray!8]
    (0,.5)--({1/3},.5)
    --({1/3},{2/3})--(0,1)--cycle;

\fill[gray!35]
    ({1/3},.5)--({3/7},{3/7})
    --(.5,.5)--({1/3},{2/3})--cycle;

\fill[gray!35]
    (.5,{1/3})--({3/7},{3/7})
    --(.5,.5)--({2/3},{1/3})--cycle;

\draw (0,0)--(1,0)--(0,1)--cycle;

\draw (.5,0)--(.5,{1/3})
    --({3/7},{3/7})--({1/3},.5)--(0,.5);
\draw (.5,{1/3})--({2/3},{1/3});
\draw ({1/3},.5)--({1/3},{2/3});
\draw ({3/7},{3/7})--(.5,.5);

\node at (.22,.20) {$h$};
\node at (.70,.15) {$x$};
\node at (.14,.70) {$y$};
\node at (.39,.54) {$e$};
\node at (.54,.39) {$f$};

\node[below left] at (0,0) {$0$};
\node[below] at (.5,0) {$1/2$};
\node[below] at (1,0) {$1$};
\node[left] at (0,.5) {$1/2$};
\node[left] at (0,1) {$1$};

\node[below] at (.5,-.09) {$s=\mu(c)$};
\node[rotate=90] at (-.17,.5) {$t=\mu(d)$};

\end{tikzpicture}
\caption{Player 1's best-reply regions.
On shared boundaries, the corresponding actions are
simultaneous best replies.}
\label{fig:br-regions}
\end{figure}

The actions $e$ and $f$ are unique best replies at $(s,t)=(2/5,3/5)$ and $(3/5,2/5)$, respectively. The actions $x$, $y$, and $h$ are unique best replies to $c$, $d$, and $a$, respectively. For a conjecture $q$ over player 1's actions, player 2's payoffs are
\[
\begin{aligned}
u_2(a,q)&=4(q_e+q_f+q_h) &\quad\text{ and }\quad u_2(b,q)&=4(q_x+q_y),\\
u_2(c,q)&=3(q_e+q_x) &\quad\text{ and }\quad u_2(d,q)&=3(q_f+q_y).
\end{aligned}
\]
The actions $a$ and $b$ are unique best replies to $e$ and $x$. The actions $c$ and $d$ are unique best replies to equal mixtures of $e,x$ and $f,y$, respectively. Thus $\rho_i(A_{-i})=A_i$ for both players, and every action is rationalizable.

As Figure~\ref{fig:br-regions} illustrates, no exact best-reply set of player 1 contains both $e$ and $x$.
To see this formally, suppose that both actions are optimal.
Then $u_1(e)\geq u_1(f)$ and $u_1(x)\geq u_1(y)$ imply $s=t$.
Equality of the payoffs of $e$ and $x$ then gives $s=t=1/3$.
Both actions yield an expected payoff of $4/3$, below the payoff of $h$.
By symmetry, no exact best-reply set contains both $f$ and $y$.

No exact best-reply set of player 2 contains both $c$ and $d$. Indeed, $u_2(a,q)+u_2(b,q)=4$, so the maximum payoff is at least 2, whereas $u_2(c,q)+u_2(d,q)\leq3$. Moreover, $c$ can be a best reply only if $q_e>0$ and $q_x>0$. If $q_e=0$, action $b$ gives strictly more than $c$ whenever $q_x>0$; if also $q_x=0$, action $c$ gives zero and is not optimal. The argument using $a$ applies when $q_x=0$. Similarly, $d$ can be a best reply only if $q_f>0$ and $q_y>0$.

Fix any $E\in\mathcal E$. Its second component misses $c$ or $d$ by the previous argument. Suppose that it misses $d$. Its first component misses at least one of $f$ and $y$. As Figure~\ref{fig:br-regions} illustrates, when $d$ is absent ($t=0$), only $x$ and $h$ can be best replies of player 1.
Indeed, both $e$ and $f$ yield strictly less than $x$, and $y$ yields strictly less than $h$.
Hence the missing member of $\{f,y\}$ cannot enter. Since a conjecture justifying $d$ must give positive probability to both $f$ and $y$, action $d$ cannot enter either. Induction shows that $d$ is absent from $\tau_2^\infty(E)$. If $c$ is absent initially, the symmetric argument uses the missing member of $\{e,x\}$ and shows that $c$ remains absent.

Thus, for every $E\in\mathcal E$, at least one of $c$ and $d$
is absent from $\tau_2^\infty(E)$. Neither $e$ nor $f$ is a best reply when either of these opponent actions is absent, so neither belongs to $\RP_1$. Finally, $x$, $y$, and $h$ are best replies to $c$, $d$, and $a$, respectively.
Each of $\{c\}$, $\{d\}$, and $\{a\}$ is an exact best-reply set for player 2.
An anchor inducing such a singleton as player 2's level-1 set makes the corresponding action of player 1 available at level 2.
\end{proof}

\section{Why smaller games cannot work}

We may restrict attention to games in which every action is rationalizable. To see this, restrict a game to $R_1^\infty\times R_2^\infty$. Against a conjecture supported on $R_{-i}^\infty$, every best reply in the original game belongs to $R_i^\infty$. Consequently, best-reply sets in the restricted game are exactly the original best-reply sets at those conjectures. Every anchor of the restricted game induces the same level-1 sets and the same cumulative recursion in the original game. If the restricted game includes all its actions in $\RP$, the original game includes all its rationalizable actions in $\RP$.

Henceforth every action is rationalizable. To rule out a counterexample, it suffices to find an $E\in\mathcal E$
such that $\tau_i^\infty(E)=A_i$ for either player.
Since $\tau^\infty(E)$ is CURB and every action is rationalizable,
we then also have $\tau_{-i}^\infty(E)=A_{-i}$.
Consequently, $\RP_i=A_i$ for both players.

In particular, suppose $A_i\setminus\{z\}$ is an exact best-reply
set and $z$ is justified by some exact best-reply set $F$ of
player $-i$.
Choosing an anchor with $E_i=A_i\setminus\{z\}$ and $E_{-i}=F$
gives $\tau_i^2(E)=A_i$.
The same argument applies if $z$ is a best reply to a pure
opponent action, since every opponent action belongs to some
exact best-reply set.

\subsection{Best-reply regions and their complex}

For each action $z\in A_i$, let
\[
P_i(z)=\{\mu\in\Delta(A_{-i}):z\in r_i(\mu)\}.
\]
These are nonempty\footnote{Recall that we assume that every action is rationalizable.} closed convex polytopes, which across all $z \in A_i$, cover the entire simplex of conjectures. Their intersections define a finite simplicial complex\footnote{A \emph{simplicial complex} on a finite vertex set $V$ is a collection of subsets of $V$ that contains every singleton and is closed under taking subsets. Its members are called faces. A face is maximal if it is not properly contained in another face.}
\[
K_i=\left\{B\subseteq A_i:\bigcap_{z\in B}P_i(z)\ne\varnothing\right\},
\]
where the empty set is included by convention. Vertices are actions, and a face is a set of actions that are best replies to the same conjecture. An edge joins two actions that can be optimal together. A triangle is filled only when its three actions can be optimal together. If all subsets are faces, the complex is a full simplex. Every maximal face is an exact best-reply set: a conjecture justifying all its actions has a best-reply set containing the face, and maximality gives equality.

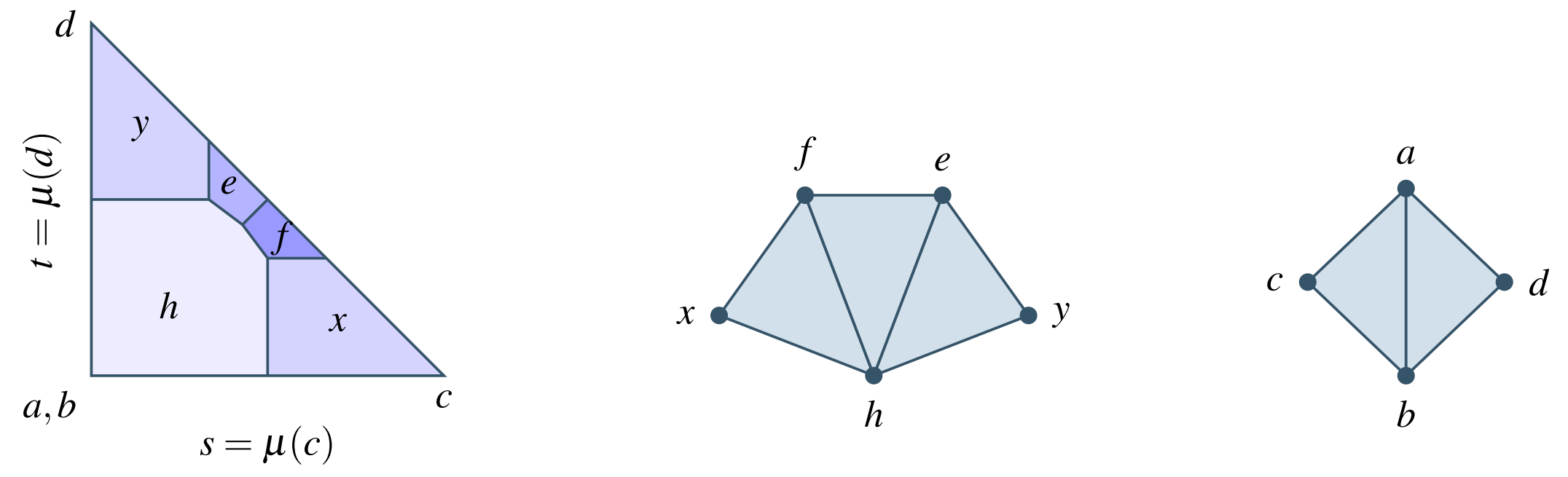
\begin{figure}[tb]
\centering
\begin{subfigure}[t]{.37\textwidth}
\centering
\begin{tikzpicture}[scale=3.2,font=\small]
\fill[blue!7] (0,0)--(.5,0)--(.5,.333333)--(.428571,.428571)--(.333333,.5)--(0,.5)--cycle;
\fill[blue!17] (.5,0)--(1,0)--(.666667,.333333)--(.5,.333333)--cycle;
\fill[blue!17] (0,.5)--(.333333,.5)--(.333333,.666667)--(0,1)--cycle;
\fill[blue!29] (.333333,.5)--(.428571,.428571)--(.5,.5)--(.333333,.666667)--cycle;
\fill[blue!40] (.5,.333333)--(.428571,.428571)--(.5,.5)--(.666667,.333333)--cycle;
\draw[edge] (0,0)--(1,0)--(0,1)--cycle;
\draw[edge] (.5,0)--(.5,.333333)--(.428571,.428571)--(.333333,.5)--(0,.5);
\draw[edge] (.5,.333333)--(.666667,.333333);
\draw[edge] (.333333,.5)--(.333333,.666667);
\draw[edge] (.428571,.428571)--(.5,.5);
\node at (.22,.2) {$h$};\node at (.70,.15) {$x$};\node at (.14,.70) {$y$};
\node at (.39,.54) {$e$};\node at (.54,.39) {$f$};
\node[below left] at (0,0) {$a,b$};\node[below] at (1,0) {$c$};\node[left] at (0,1) {$d$};
\node[below] at (.5,-.10) {$s=\mu(c)$};\node[rotate=90] at (-.14,.5) {$t=\mu(d)$};
\end{tikzpicture}
\caption{Player 1's best-reply regions.}
\end{subfigure}\hfill
\begin{subfigure}[t]{.33\textwidth}
\centering
\begin{tikzpicture}[scale=.78,font=\small]
\coordinate (h) at (0,0);\coordinate (x) at (-1.8,.7);\coordinate (f) at (-.8,2.1);\coordinate (e) at (.8,2.1);\coordinate (y) at (1.8,.7);
\path[face] (h)--(x)--(f)--cycle;
\path[face] (h)--(f)--(e)--cycle;
\path[face] (h)--(e)--(y)--cycle;
\foreach \v/\pos in {h/below,x/left,f/above,e/above,y/right}{\node[vertex,label=\pos:{$\v$}] at (\v) {};}
\path[use as bounding box] (-2.2,-1.2) rectangle (2.2,3.3);
\end{tikzpicture}
\caption{The complex $K_1$.}
\end{subfigure}\hfill
\begin{subfigure}[t]{.25\textwidth}
\centering
\begin{tikzpicture}[scale=.85,font=\small]
\coordinate (a) at (0,2);\coordinate (b) at (0,0);\coordinate (c) at (-1.05,1);\coordinate (d) at (1.05,1);
\path[face] (a)--(b)--(c)--cycle;\path[face] (a)--(b)--(d)--cycle;
\foreach \v/\pos in {a/above,b/below,c/left,d/right}{\node[vertex,label=\pos:{$\v$}] at (\v) {};}
\path[use as bounding box] (-1.6,-1.1) rectangle (1.6,3.0);
\end{tikzpicture}
\caption{The complex $K_2$.}
\end{subfigure}
\caption{Best-reply regions and simplicial complexes in the $5\times4$ example.
Panel (a) reproduces Figure~\ref{fig:br-regions}.
In panel (b), the maximal faces are $\{h,x,f\}$, $\{h,f,e\}$, and $\{h,e,y\}$.
In panel (c), they are $\{a,b,c\}$ and $\{a,b,d\}$.
The Euler characteristics are $5-7+3=1$ and $4-5+2=1$, respectively.}
\label{fig:example}
\end{figure}

Figure~\ref{fig:example} illustrates the construction for the $5\times4$ example.
Panel (a) reproduces the best-reply regions from Figure~\ref{fig:br-regions}, while panels (b) and (c) depict the simplicial complexes of players 1 and 2, respectively.\footnote{The positions of the vertices in the latter two panels have no payoff interpretation.}
Each face corresponds to a set of actions whose best-reply regions
have a nonempty intersection.

\begin{lemma}\label{lem:complex}
The complex $K_i$ is connected and has Euler characteristic one.
\end{lemma}

\begin{proof}
The regions $P_i(z)$ form a finite closed cover of the
simplex of conjectures.
Every nonempty intersection of these regions is convex
and hence contractible.
By the nerve theorem \citep[Theorem 10.7]{Bjorner1995},
the simplicial complex $K_i$ is homotopy equivalent to
the union of these regions, which is $\Delta(A_{-i})$.
Since the simplex is contractible, so is $K_i$.
In particular, $K_i$ is connected and has Euler
characteristic one.\footnote{For other applications of the nerve theorem in economic
theory, see \citet{Lauwers2000} in social choice theory and
\citet{KushnirLokutsievskiy2021} in mechanism design.}
\end{proof}

For a complex of dimension at most three,\footnote{The dimension of a
simplicial complex is the cardinality of its largest face minus one.
In our setting, this is the maximum number of actions that can be
best replies to the same conjecture, minus one.}
the Euler characteristic is the number of vertices minus the number
of edges plus the number of triangles minus the number of tetrahedra. Since $K_i$ has dimension at most $|A_i|-1$, Lemma~\ref{lem:complex}
restricts the possible configurations of best-reply sets when
a player has few actions. We next record two implications of the graph formed by the vertices and edges of $K_i$.

\begin{lemma}\label{lem:tree}
If a vertex of the graph of $K_i$ has exactly one neighbor, its action is a best reply to a pure opponent action.
If the graph of $K_i$ is a tree,\footnote{A tree is a connected graph without cycles. Equivalently, there is a unique simple path between any two distinct vertices. A vertex with exactly one neighbor is called a leaf.}
every action of player $i$ is justified by an opponent exact best-reply set and therefore belongs to $\RP_i$.
\end{lemma}

\begin{proof}
Let $z$ have the unique neighbor $p$, and choose $\mu\in P_i(z)$. Since $u_i(z,\mu)\geq u_i(p,\mu)$, there is a pure opponent action $b$ with $u_i(z,b)\geq u_i(p,b)$. Along the segment from $\mu$ to the degenerate conjecture assigning probability one to $b$, action $z$ gives at least as much as $p$. The best-reply regions other than $P_i(p)$ therefore still cover this segment: whenever $p$ is optimal, $z$ is optimal too.

The intersection graph of the nonempty members of a finite closed cover of a connected set is connected. Indeed, a partition of that graph into two nonempty parts with no edge between them would separate the set into two disjoint nonempty closed subsets. Intersect each best-reply region other than $P_i(p)$ with
the segment and consider the intersection graph of the
nonempty resulting sets.
This graph is connected, contains $z$, and is a subgraph
of the graph obtained from $K_i$ by removing $p$ and its
incident edges.
Since $z$ has no other neighbor, this graph consists
only of $z$.
Hence $z$ is optimal throughout the segment, including at $b$. But $z$ is isolated in the latter graph. Hence $z$ is optimal throughout the segment, including at $b$.

Now suppose the graph of $K_i$ is a tree. Let $Y$ be the union of $\Delta(F)$ over all maximal faces $F$ of $K_{-i}$. The set $Y$ is the geometric realization of $K_{-i}$ in
$\Delta(A_{-i})$ and is therefore connected by
Lemma~\ref{lem:complex}.
Since every opponent action is rationalizable, $Y$
contains every degenerate conjecture in $\Delta(A_{-i})$. The nonempty sets $P_i(z)\cap Y$ form a finite closed cover of $Y$, so the actions justified on $Y$ form a connected subgraph of the graph of $K_i$. By the first part, this subgraph contains every leaf. A connected subgraph of a finite tree containing every leaf contains the whole tree. Thus every action is justified by some $\Delta(F)$, where $F$ is an opponent's exact best-reply set. An anchor that induces $F$ as the opponent's level-1 set makes that action available at level 2. The claim is immediate if the tree has only one vertex.
\end{proof}

\begin{figure}[htbp]
\centering
\begin{subfigure}[b]{.48\textwidth}
\centering
\begin{tikzpicture}[scale=5]

\fill[blue!15]
    (0,0)--(.25,0)--(.25,.75)--(0,1)--cycle;
\fill[gray!15]
    (.25,0)--(.75,0)--(.75,.25)--(.25,.75)--cycle;
\fill[orange!20]
    (.75,0)--(1,0)--(.75,.25)--cycle;

\draw (0,0)--(1,0)--(0,1)--cycle;
\draw (.25,0)--(.25,.75);
\draw (.75,0)--(.75,.25);

\draw[very thick] (0,0)--(0,1)--(1,0);
\node[above right] at (.5,.5) {$Y$};

\draw[dashed,thick,->] (.25,.4)--(0,0);
\fill (.25,.4) circle (1pt);
\node[right] at (.26,.4) {$\mu$};

\node at (.12,.55) {$P_1(z)$};
\node at (.48,.24) {$P_1(p)$};
\node at (.835,.05) {$P_1(q)$};

\node[below left] at (0,0) {$b$};
\node[below right] at (1,0) {$c$};
\node[above] at (0,1) {$d$};

\node[below] at (.25,0) {$1/4$};
\node[below] at (.75,0) {$3/4$};
\node[below] at (.5,-.15) {$\mu(c)$};

\end{tikzpicture}
\caption{Best-reply regions.}
\end{subfigure}
\hfill
\begin{subfigure}[b]{.48\textwidth}
\centering
\begin{tikzpicture}[scale=2]

\draw[thick] (0,0)--(1,0)--(2,0);

\fill[blue!15] (0,0) circle (3pt);
\fill[gray!15] (1,0) circle (3pt);
\fill[orange!20] (2,0) circle (3pt);

\foreach \x/\a in {0/z,1/p,2/q}{
    \draw (\x,0) circle (3pt);
    \node[above=4pt] at (\x,0) {$\a$};
}

\node[below=5pt] at (0,0) {\small Leaf};
\node[below=5pt] at (2,0) {\small Leaf};

\end{tikzpicture}
\caption{The complex $K_1$.}
\end{subfigure}

\caption{Illustration of Lemma~\ref{lem:tree}.
The thick boundary in panel (a) represents $Y$.
The dashed segment illustrates the first part of the lemma.
In panel (b), every connected subgraph containing both
leaves also contains $p$.}
\label{fig:tree-example}
\end{figure}
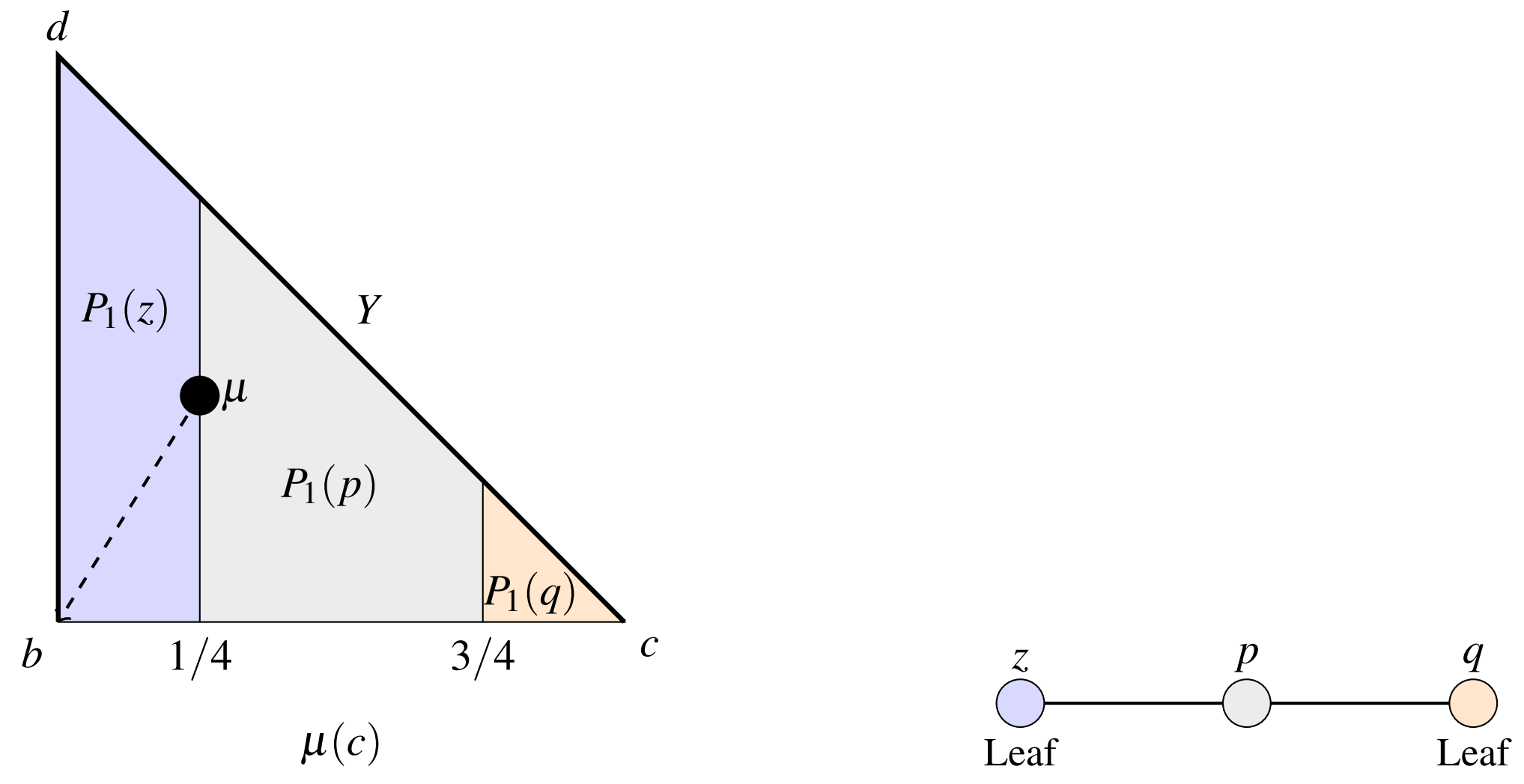

To illustrate Lemma~\ref{lem:tree}, consider the game
\[
\begin{array}{c|ccc}
  & b & c & d\\ \hline
z & 4,4 & 0,0 & 4,3\\
p & 3,0 & 3,0 & 3,3\\
q & 0,0 & 4,4 & 0,3
\end{array}
\]
Figure~\ref{fig:tree-example} depicts player 1's best-reply
regions and the corresponding simplicial complex.
The leaves $z$ and $q$ are best replies to pure opponent
actions, whereas $p$ is not.
The maximal faces of $K_2$ are $\{b,d\}$ and $\{c,d\}$.
Their corresponding simplexes form the connected set $Y$
depicted in panel (a).
Since $Y$ intersects the best-reply regions of both leaves,
it must also intersect the best-reply region of $p$.

\subsection{A restriction with four actions}

The remaining argument uses \citet[Proposition 1]{Weinstein2020}. His result implies that if two actions cannot be best replies to the same conjecture, some mixed action strictly dominates a mixture of them. The dominating and dominated mixtures can be chosen to have disjoint supports. When the player has only four actions, this has a useful consequence.

\begin{lemma}\label{lem:four}
Suppose a player has four actions $p,q,x,y$. Assume that $x$ and $y$ are each best replies to some conjecture, but are never best replies to the same conjecture. There cannot be two distinct opponent actions $c,d$ such that every conjecture justifying either $x$ or $y$ assigns positive probability to both $c$ and $d$.
\end{lemma}

\begin{proof}
By Weinstein's result, there are a mixed action $\sigma$ supported on $\{p,q\}$ and $\lambda\in(0,1)$ such that
\[
(\forall b\in A_{-i})\quad
u_i(\sigma,b)>\lambda u_i(x,b)+(1-\lambda)u_i(y,b).
\]
To obtain this form, note that the dominated mixture must put positive weight on both $x$ and $y$: neither action can itself be strictly dominated, since each is justifiable. Disjointness then confines the dominating mixture to $\{p,q\}$.

If $x$ gives at least as much as both $p$ and $q$, it gives at least as much as $\sigma$. The strict inequality above then implies that $x$ gives strictly more than $y$. Thus $x$ is a best reply precisely when its payoff is at least the payoffs of $p$ and $q$. The analogous statement holds for $y$.

Each of $x$ and $y$ can therefore be justified by a conjecture with at most two opponent actions in its support. To verify this for $x$, consider the convex hull in $\mathbb R^2$ of
\[
\left\{\bigl(u_i(x,b)-u_i(p,b),\ u_i(x,b)-u_i(q,b)\bigr):b\in A_{-i}\right\}.
\]
Panel (a) of Figure~\ref{fig:four-proof} illustrates the
following argument.
This hull meets the nonnegative quadrant because $x$ is justifiable. If the hull is two-dimensional, move from such a point in direction $(1,1)$ until reaching its boundary. The resulting point is still nonnegative and lies on an edge, so it is a convex combination of at most two of the displayed points. If the hull has smaller dimension, at most two points suffice directly. The corresponding conjecture justifies $x$.

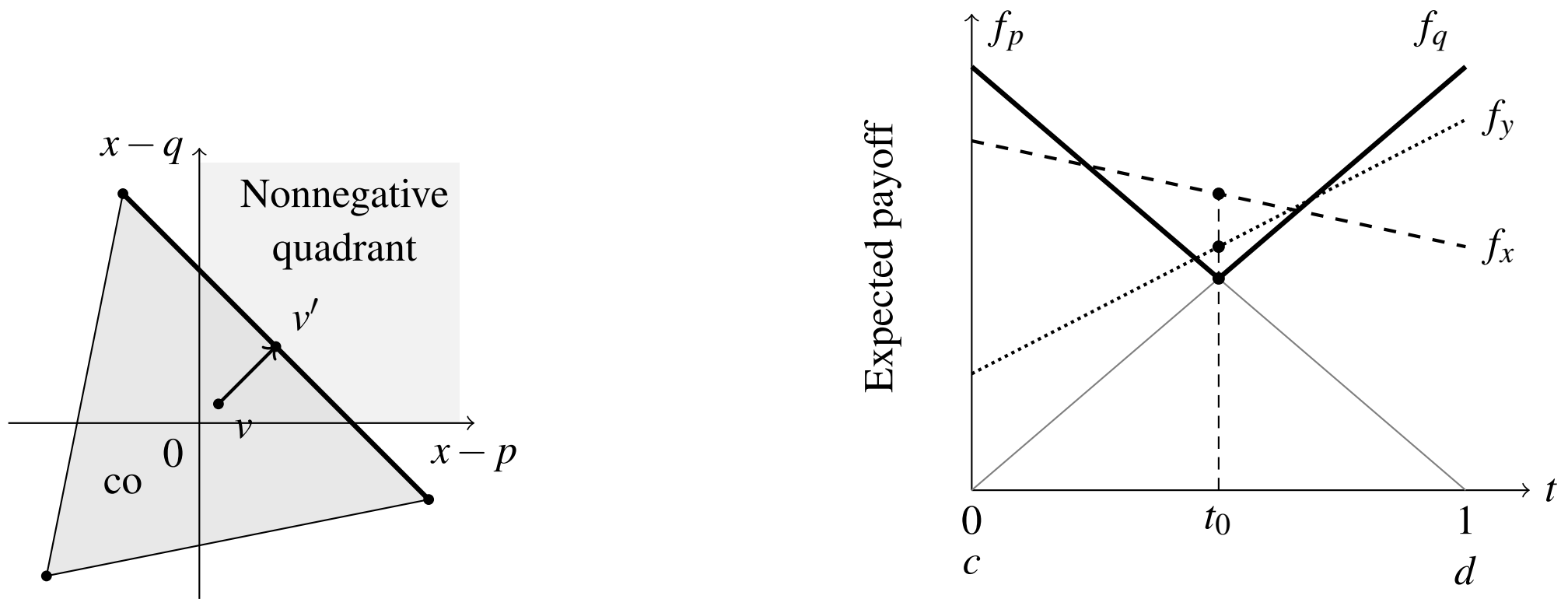
\begin{figure}[htbp]
\centering
\begin{subfigure}[b]{.48\textwidth}
\centering
\begin{tikzpicture}[scale=.65,font=\small]

\fill[gray!10] (0,0) rectangle (3.4,3.4);

\fill[gray!25,fill opacity=.7]
    (-2,-2)--(-1,3)--(3,-1)--cycle;
\draw
    (-2,-2)--(-1,3)--(3,-1)--cycle;

\draw[->] (-2.5,0)--(3.6,0)
    node[below] {$x-p$};
\draw[->] (0,-2.3)--(0,3.6)
    node[left] {$x-q$};

\foreach \a/\b in {-2/-2,-1/3,3/-1}{
    \fill (\a,\b) circle (2pt);
}

\draw[very thick] (-1,3)--(3,-1);

\draw[->,thick] (.25,.25)--(1,1);
\fill (.25,.25) circle (2pt);
\fill (1,1) circle (2pt);

\node[below right] at (.25,.25) {$v$};
\node[above right] at (1,1) {$v'$};

\node[align=center] at (1.9,2.6)
    {Nonnegative\\quadrant};
\node[align=center] at (-1,-.8)
    {$\operatorname{co}$};

\node[below left] at (0,0) {$0$};

\end{tikzpicture}
\caption{Reducing the support of a conjecture.}
\end{subfigure}
\hfill
\begin{subfigure}[b]{.48\textwidth}
\centering
\begin{tikzpicture}[x=4.2cm,y=1.8cm,font=\small]

\draw[->] (0,0)--(1.13,0) node[right] {$t$};
\draw[->] (0,0)--(0,2.25);

\draw[gray] (0,2)--(1,0);
\draw[gray] (0,0)--(1,2);

\draw[very thick] (0,2)--(.5,1)--(1,2);

\draw[thick,dashed] (0,1.65)--(1,1.15);
\draw[thick,densely dotted] (0,.55)--(1,1.75);

\draw[densely dashed] (.5,0)--(.5,1.4);

\fill (.5,1) circle (1.5pt);
\fill (.5,1.4) circle (1.5pt);
\fill (.5,1.15) circle (1.5pt);

\node[above right] at (0,2) {$f_p$};
\node[above left] at (1,2) {$f_q$};
\node[right] at (1,1.15) {$f_x$};
\node[right] at (1,1.75) {$f_y$};

\node[below] at (0,0) {$0$};
\node[below] at (.5,0) {$t_0$};
\node[below] at (1,0) {$1$};
\node[below=12pt] at (0,0) {$c$};
\node[below=12pt] at (1,0) {$d$};

\node[rotate=90] at (-.18,1.1)
    {Expected payoff};

\end{tikzpicture}
\caption{Payoff functions on $\Delta(\{c,d\})$.}
\end{subfigure}

\caption{Illustration of the proof of Lemma~\ref{lem:four}.
In panel (a), the axes measure the expected payoff differences
between $x$ and $p$ and between $x$ and $q$, respectively.
Moving from $v$ in direction $(1,1)$ yields a boundary
point $v'$ that is a convex combination of two extreme points.
Panel (b) illustrates the configuration ruled out by strict
dominance. Both $f_x$ and $f_y$ reach or exceed the common
payoff of $p$ and $q$ at $t_0$.}
\label{fig:four-proof}
\end{figure}

Suppose, toward a contradiction, that every conjecture justifying
either $x$ or $y$ assigns positive probability to both $c$ and $d$.
By the preceding argument, each of $x$ and $y$ is then justified
by some conjecture with support $\{c,d\}$, that is, by some
conjecture in the relative interior of $\Delta(\{c,d\})$. Write $t\in[0,1]$ for the probability assigned to $d$.
The expected payoff of action $z$ is then
\[
f_z(t)=(1-t)u_i(z,c)+tu_i(z,d).
\] At both endpoints, $f_x$ and $f_y$ are strictly below $\max\{f_p,f_q\}$, because neither action is optimal there. In the interior, each reaches or exceeds this maximum somewhere.

The lines $f_p$ and $f_q$ must therefore cross at a unique interior point $t_0$. Panel (b) of Figure~\ref{fig:four-proof} illustrates this
configuration. Otherwise their maximum would be affine on the whole edge, and an affine function strictly below it at both endpoints would be strictly below it everywhere. For each $z\in\{x,y\}$, the function $f_z-\max\{f_p,f_q\}$ is affine on each side of $t_0$, negative at both endpoints, and nonnegative somewhere. Consequently,
\[
f_x(t_0)\geq f_p(t_0)=f_q(t_0)
\quad\text{ and }\quad
f_y(t_0)\geq f_p(t_0)=f_q(t_0).
\]
At $t_0$, every mixture of $p$ and $q$ gives their common payoff. It cannot strictly dominate the indicated mixture of $x$ and $y$, a contradiction.
\end{proof}

\subsection{The lower bound in Theorem~\ref{thm:minimum}}

\begin{proof}
By the reduction above, assume that every action is rationalizable. First suppose that one player has at most three actions. By Lemma~\ref{lem:complex}, that player's complex is either a full simplex or a path on three vertices. In the first case, the whole action set is an exact best-reply set. Choosing an anchor with $E_i=A_i$ gives
$\tau^\infty(E)=A_1\times A_2$. In the second case, the complex has two maximal faces,
each containing two actions and omitting a leaf of the graph.
By Lemma~\ref{lem:tree}, the omitted action is a best reply
to a pure opponent action. We can therefore choose an anchor under which the maximal edge
is player $i$'s level-1 set and the omitted action is available
at level 2.
This gives $\tau_i^2(E)=A_i$ and hence
$\tau^\infty(E)=A_1\times A_2$. Thus a counterexample requires at least four actions for each player.

It remains to rule out $4\times4$.
Figure~\ref{fig:four} depicts, up to relabeling, all connected
simplicial complexes on four vertices with Euler characteristic one.
If the complex contains a tetrahedron, it is the full simplex.
Otherwise, writing $m$ for the number of edges and $t$ for
the number of triangles, the Euler characteristic condition gives
\[
4-m+t=1.
\]
Since there are at most six edges, $t$ can take only the
values $0,1,2,3$.
If $t=0$, the connected graph has three edges and is thus
either a path or a star.
If $t=1$, connectedness requires the fourth edge to join
the remaining vertex to the triangle.
If $t=2$, the two triangles necessarily share an edge
and together account for all five edges.
If $t=3$, the complex consists of three faces of a
tetrahedron, which necessarily share a common vertex.
Four triangles without the tetrahedron would give an Euler
characteristic of two, contradicting Lemma~\ref{lem:complex}.
This exhausts all possibilities.

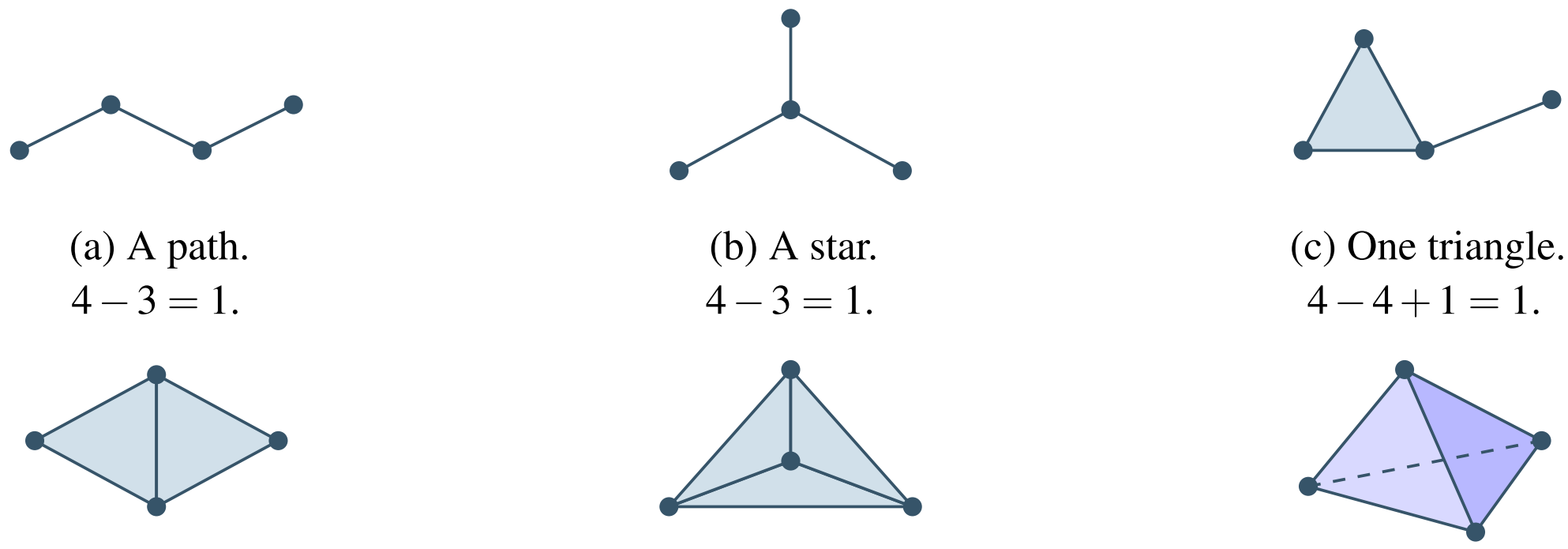
\begin{figure}[tb]
\centering
\begin{subfigure}[t]{.31\textwidth}\centering
\begin{tikzpicture}[scale=.85]
\draw[edge] (-1.35,0)--(-.45,.45)--(.45,0)--(1.35,.45);
\foreach \x/\y in {-1.35/0,-.45/.45,.45/0,1.35/.45}{\node[vertex] at (\x,\y) {};}
\path[use as bounding box] (-1.55,-.45) rectangle (1.55,1.45);
\end{tikzpicture}
\caption{A path.\\ $4-3=1$.}
\end{subfigure}\hfill
\begin{subfigure}[t]{.31\textwidth}\centering
\begin{tikzpicture}[scale=.85]
\draw[edge] (0,.4)--(-1.1,-.2);\draw[edge] (0,.4)--(1.1,-.2);\draw[edge] (0,.4)--(0,1.3);
\foreach \x/\y in {0/.4,-1.1/-.2,1.1/-.2,0/1.3}{\node[vertex] at (\x,\y) {};}
\path[use as bounding box] (-1.55,-.45) rectangle (1.55,1.45);
\end{tikzpicture}
\caption{A star.\\ $4-3=1$.}
\end{subfigure}\hfill
\begin{subfigure}[t]{.31\textwidth}\centering
\begin{tikzpicture}[scale=.85]
\path[face] (-1.2,0)--(0,0)--(-.6,1.1)--cycle;\draw[edge] (0,0)--(1.25,.5);
\foreach \x/\y in {-1.2/0,0/0,-.6/1.1,1.25/.5}{\node[vertex] at (\x,\y) {};}
\path[use as bounding box] (-1.55,-.45) rectangle (1.55,1.45);
\end{tikzpicture}
\caption{One triangle.\\ $4-4+1=1$.}
\end{subfigure}

\medskip
\begin{subfigure}[t]{.31\textwidth}\centering
\begin{tikzpicture}[scale=.85]
\path[face] (0,0)--(0,1.3)--(-1.2,.65)--cycle;\path[face] (0,0)--(0,1.3)--(1.2,.65)--cycle;
\foreach \x/\y in {0/0,0/1.3,-1.2/.65,1.2/.65}{\node[vertex] at (\x,\y) {};}
\path[use as bounding box] (-1.55,-.45) rectangle (1.55,1.45);
\end{tikzpicture}
\caption{Two triangles.\\ $4-5+2=1$.}
\end{subfigure}\hfill
\begin{subfigure}[t]{.31\textwidth}\centering
\begin{tikzpicture}[scale=.85]
\path[face] (-1.2,0)--(1.2,0)--(0,.45)--cycle;
\path[face] (-1.2,0)--(0,1.35)--(0,.45)--cycle;
\path[face] (1.2,0)--(0,1.35)--(0,.45)--cycle;
\foreach \x/\y in {-1.2/0,1.2/0,0/1.35,0/.45}{\node[vertex] at (\x,\y) {};}
\path[use as bounding box] (-1.55,-.45) rectangle (1.55,1.45);
\end{tikzpicture}
\caption{Three triangles.\\ $4-6+3=1$.}
\end{subfigure}\hfill
\begin{subfigure}[t]{.31\textwidth}\centering
\begin{tikzpicture}[scale=.85]
\coordinate (a) at (-1.15,.2);\coordinate (b) at (.5,-.25);\coordinate (c) at (1.15,.65);\coordinate (d) at (-.2,1.35);
\path[fill=blue!15] (a)--(b)--(d)--cycle;
\path[fill=blue!28] (b)--(c)--(d)--cycle;
\draw[edge,dashed] (a)--(c);
\draw[edge] (a)--(b)--(c)--(d)--(a);\draw[edge] (b)--(d);
\foreach \v in {a,b,c,d}{\node[vertex] at (\v) {};}
\path[use as bounding box] (-1.55,-.45) rectangle (1.55,1.45);
\end{tikzpicture}
\caption{A tetrahedron.\\ $4-6+4-1=1$.}
\end{subfigure}
\caption{The possible best-reply complexes on four actions, up to relabeling. Shaded triangles are faces. Panel (e) has three triangles meeting at the central vertex; the outer triangle is not a face. Panel (f) represents a three-dimensional simplex, including its interior.}
\label{fig:four}
\end{figure}

Suppose, toward a contradiction, that some rationalizable action
is excluded from $\RP_i$ for either player.
Then $\tau^\infty(E)\ne A_1\times A_2$ for every $E\in\mathcal E$. Neither player's complex can be a full simplex.
As in the case of at most three actions, we could otherwise
choose an anchor with $E_i=A_i$, giving
$\tau^\infty(E)=A_1\times A_2$. A triangle with an attached edge is also impossible.
Its maximal triangle is an exact best-reply set containing
all but one action, and the omitted action is a leaf.
By Lemma~\ref{lem:tree}, the latter is a best reply to a
pure opponent action.
We can therefore choose an anchor under which the three
actions of the triangle are available at level 1 and the
remaining action is available at level 2.
This gives $\tau_i^2(E)=A_i$ and hence
$\tau^\infty(E)=A_1\times A_2$. Nor can either complex consist of three faces of a tetrahedron.
These three faces share one vertex.
For each of the other three vertices, there is a maximal
face containing all but that vertex.
If any of these three actions were a best reply to a pure
opponent action, we could again find an $E\in\mathcal E$
such that $\tau^\infty(E)=A_1\times A_2$.
Otherwise, the action corresponding to the common vertex
would be the unique best reply to every pure opponent action.
It would therefore strictly dominate all three other actions,
contradicting rationalizability.

Each player's complex is therefore a tree or consists of two triangles sharing an edge. Relabel players so that player 1 has an excluded action. Lemma~\ref{lem:tree} rules out a tree for player 1. Write player 1's two maximal faces as $F=\{p,q,x\}$ and $F'=\{p,q,y\}$.

Suppose player 2's complex were a tree. By Lemma~\ref{lem:tree}, each of player 2's actions would be justified by $F$ or $F'$, so
\[
\rho_2(F)\cup\rho_2(F')=A_2
\quad\text{ and }\quad
\rho_2(F)\cap\rho_2(F')\supseteq\rho_2(\{p,q\})\ne\varnothing.
\]
Since $|A_2|=4$, at least one of $\rho_2(F)$ and
$\rho_2(F')$ contains at least three actions.
Without loss of generality, suppose this is $\rho_2(F)$.
Choose an anchor with $E_1=F$.
If $\rho_2(F)$ omits an action, choose $E_2$ to contain
that action.
We then have $\tau_2^2(E)=A_2$ and hence
$\tau^\infty(E)=A_1\times A_2$.
Thus player 2's complex also consists of two triangles
sharing an edge, say $\{a,b,c\}$ and $\{a,b,d\}$.

Suppose $x$ were justified by either maximal face of $K_2$.
Choosing that face as $E_2$ and $\{p,q,y\}$ as $E_1$
would give $\tau_1^2(E)=A_1$ and hence
$\tau^\infty(E)=A_1\times A_2$.
The analogous argument, with $E_1=\{p,q,x\}$, applies to $y$. Hence every conjecture justifying either $x$ or $y$ must assign positive probability to both $c$ and $d$. But $x$ and $y$ are individually justifiable and are never best replies to the same conjecture, since $\{x,y\}$ is not an edge of $K_1$. Lemma~\ref{lem:four} gives a contradiction. This rules out $4\times4$ and proves the lower bound. Proposition~\ref{prop:example} proves attainment.
\end{proof}

\bibliographystyle{ecta}
\bibliography{bib}

@unpublished{LiuZiegler2026,
  author = {Liu, Shuige and Ziegler, Gabriel},
  title = {Reasoning About Bounded Reasoning},
  year = {2026},
  note = {Working paper, arXiv:2506.19737},
  url = {https://arxiv.org/abs/2506.19737}
}

@article{BrandenburgerFriedenbergKeisler2008,
  author  = {Brandenburger, Adam and Friedenberg, Amanda
             and Keisler, H. Jerome},
  title   = {Admissibility in Games},
  journal = {Econometrica},
  year    = {2008},
  volume  = {76},
  number  = {2},
  pages   = {307--352},
  doi     = {10.1111/j.1468-0262.2008.00835.x}
}

@article{Lauwers2000,
  author  = {Lauwers, Luc},
  title   = {Topological Social Choice},
  journal = {Mathematical Social Sciences},
  year    = {2000},
  volume  = {40},
  number  = {1},
  pages   = {1--39},
  doi     = {10.1016/S0165-4896(99)00036-0}
}

@article{KushnirLokutsievskiy2021,
  author  = {Kushnir, Alexey I. and Lokutsievskiy, Lev V.},
  title   = {When Is a Monotone Function Cyclically Monotone?},
  journal = {Theoretical Economics},
  year    = {2021},
  volume  = {16},
  number  = {3},
  pages   = {853--879},
  doi     = {10.3982/TE4305}
}

@phdthesis{Egan2008,
  author = {Egan, Sarah Jane},
  title  = {{Nash Equilibria in Games and Simplicial Complexes}},
  school = {University of Bath},
  year   = {2008},
  type   = {Ph.D. thesis}
}

@article{Weinstein2020,
  author = {Weinstein, Jonathan},
  title = {Best-Reply Sets},
  journal = {Economic Theory Bulletin},
  year = {2020},
  volume = {8},
  pages = {105--112},
  doi = {10.1007/s40505-019-00169-1}
}

@unpublished{BattigalliCatoniniDeVito2025,
  author = {Battigalli, Pierpaolo and Catonini, Emiliano and De Vito, Nicodemo},
  title = {Game Theory: Analysis of Strategic Thinking},
  year = {2025},
  note = {Book in preparation},
  url = {https://didattica.unibocconi.it/mypage/upload/48808_20260526_055225_TEXTBOOOKGT-AST_DEC2025PARTI.PDF}
}

@unpublished{BrandenburgerFriedenbergKneeland2025,
  author = {Brandenburger, Adam and Friedenberg, Amanda and Kneeland, Terri},
  title = {From Anchors to Hierarchies: Identifying Levels of Reasoning in Games},
  year = {2025},
  note = {mimeo},
  url = {https://www.amandafriedenberg.org/s/levelk.pdf}
}

@unpublished{HalevyHoelzemannKneeland2025,
  author = {Halevy, Yoram and Hoelzemann, Johannes C. and Kneeland, Terri},
  title = {Magic Mirror on the Wall, Who Is the Smartest One of All?},
  year = {2025},
  note = {mimeo},
  url = {https://www.johanneshoelzemann.com/papers/magic_mirror.pdf}
}

@article{StahlWilson1995,
  author = {Stahl, Dale O. and Wilson, Paul W.},
  title = {On Players' Models of Other Players: Theory and Experimental Evidence},
  journal = {Games and Economic Behavior},
  year = {1995},
  volume = {10},
  number = {1},
  pages = {218--254},
  doi = {10.1006/game.1995.1031}
}

@article{Nagel1995,
  author = {Nagel, Rosemarie},
  title = {Unraveling in Guessing Games: An Experimental Study},
  journal = {American Economic Review},
  year = {1995},
  volume = {85},
  number = {5},
  pages = {1313--1326}
}

@article{CamererHoChong2004,
  author = {Camerer, Colin F. and Ho, Teck-Hua and Chong, Juin-Kuan},
  title = {A Cognitive Hierarchy Model of Games},
  journal = {Quarterly Journal of Economics},
  year = {2004},
  volume = {119},
  number = {3},
  pages = {861--898},
  doi = {10.1162/0033553041502225}
}

@article{BasuWeibull1991,
  author = {Basu, Kaushik and Weibull, J{\"o}rgen W.},
  title = {Strategy Subsets Closed under Rational Behavior},
  journal = {Economics Letters},
  year = {1991},
  volume = {36},
  number = {2},
  pages = {141--146},
  doi = {10.1016/0165-1765(91)90179-O}
}

@incollection{Bjorner1995,
  author = {Bj{\"o}rner, Anders},
  title = {Topological Methods},
  booktitle = {Handbook of Combinatorics},
  editor = {Graham, Ronald L. and Gr{\"o}tschel, Martin and Lov{\'a}sz, L{\'a}szl{\'o}},
  publisher = {Elsevier},
  address = {Amsterdam},
  year = {1995},
  volume = {2},
  pages = {1819--1872}
}

@incollection{vonStengel2002,
  author = {von Stengel, Bernhard},
  title = {Computing Equilibria for Two-Person Games},
  booktitle = {Handbook of Game Theory with Economic Applications},
  editor = {Aumann, Robert J. and Hart, Sergiu},
  publisher = {North-Holland},
  address = {Amsterdam},
  year = {2002},
  volume = {3},
  pages = {1723--1759},
  doi = {10.1016/S1574-0005(02)03008-4}
}
\end{document}